\documentclass{aa} 
\usepackage{graphicx}
\usepackage{xspace}
\usepackage{natbib}
\bibpunct{(}{)}{;}{a}{}{,} 
\newcommand{\Al}{$^{26}$Al\xspace}

\newcommand{\degree}{$^{\circ}$}

\newcommand{\Msol}{M\ensuremath{_\odot}\xspace}

\begin{document}

   \title{Radioactive $^{26}$Al from the Scorpius-Centaurus Association}

\author{R. Diehl\inst{1}, M.G. Lang\inst{1}, P. Martin\inst{1}, H. Ohlendorf\inst{1}, Th. Preibisch\inst{2}, R. Voss\inst{1,4}, P. Jean\inst{5}, J.-P. Roques\inst{5}, P. von Ballmoos\inst{5}, and W. Wang\inst{3}}

\institute{Max-Planck-Institut f\"{u}r extraterrestrische Physik, Postfach 1312, 85741 Garching, Germany
\and
Universit\"atssternwarte, Scheinerstr. 1, 81679 M\"unchen, Germany
\and
National Astronomical Observatories, Chinese Academy of Sciences, Beijing 100012, China
\and
Excellence-Cluster "Origin and Structure of the Universe" , 85748 Garching, Germany
\and
Centre d'Etude Spatiale des Rayonnements, B.P.N$\circ$ 4346, 31028 Toulouse Cedex 4, France
\and
Astrophysikalisches Institut der Universit\"at, Potsdam, Germany
}

   \date{Received 20 Feb, 2010; revised 15 Jul, 2010; accepted 18 Jul, 2010}

\authorrunning{R.~Diehl {\it et al.}}
  \abstract
   {The Scorpius-Centaurus association is the most-nearby group of massive and young stars. As nuclear-fusion products are ejected by massive stars and supernovae into the surrounding interstellar medium, the search for characteristic $\gamma$-rays from radioactivity is one way to probe the history of activity of such nearby massive stars on a My time scale through their nucleosynthesis. \Al decays with a radioactivity lifetime $\tau\sim$1~My, 1809~keV $\gamma$-rays from its decay can be measured with current $\gamma$-ray telescopes.}
   {We aim to identify nucleosynthesis ejecta from the youngest subgroup of Sco-Cen stars, and interpret their location and bulk motion from \Al observations with INTEGRAL's $\gamma$-ray spectrometer SPI.}
   {Following earlier \Al $\gamma$-ray mapping with NASA's Compton observatory, we test spatial emission skymaps of \Al for a component which could be attributed to ejecta from massive stars in the Scorpius-Centaurus group of stars. Such a model fit of spatial distributions for large-scale and local components is able to discriminate \Al emission associated with Scorpius-Centaurus, in spite of the strong underlying nucleosynthesis signal from the Galaxy at large. }
   {We find an \Al $\gamma$-ray signal above 5$\sigma$ significance, which we associate with the locations of stars of the Sco-Cen group. The observed flux of 6~10$^{-5}$ph~cm$^{-2}$s$^{-1}$ corresponds to $\sim$1.1~10$^{-4}$~\Msol of \Al. This traces the nucleosynthesis ejecta of several massive stars within the past several million years. }
   {We confirm through direct detection of radioactive \Al the recent ejection of massive-star nucleosynthesis products from the Sco-Cen association.  Its  youngest subgroup in Upper Scorpius appears to dominate \Al contributions from this association. Our \Al signal can be interpreted as a measure of the age and richness of this youngest subgroup. 
   We also estimate a kinematic imprint of these nearby massive-star ejecta from the bulk motion of \Al and compare this to other indications of Scorpius-Centaurus massive-star activity .}

   \keywords{nucleosynthesis --
                massive stars and supernovae
               }

   \maketitle
%

\section{Introduction}

\begin{figure}   
   \centering
   \includegraphics[width=0.4\textwidth]{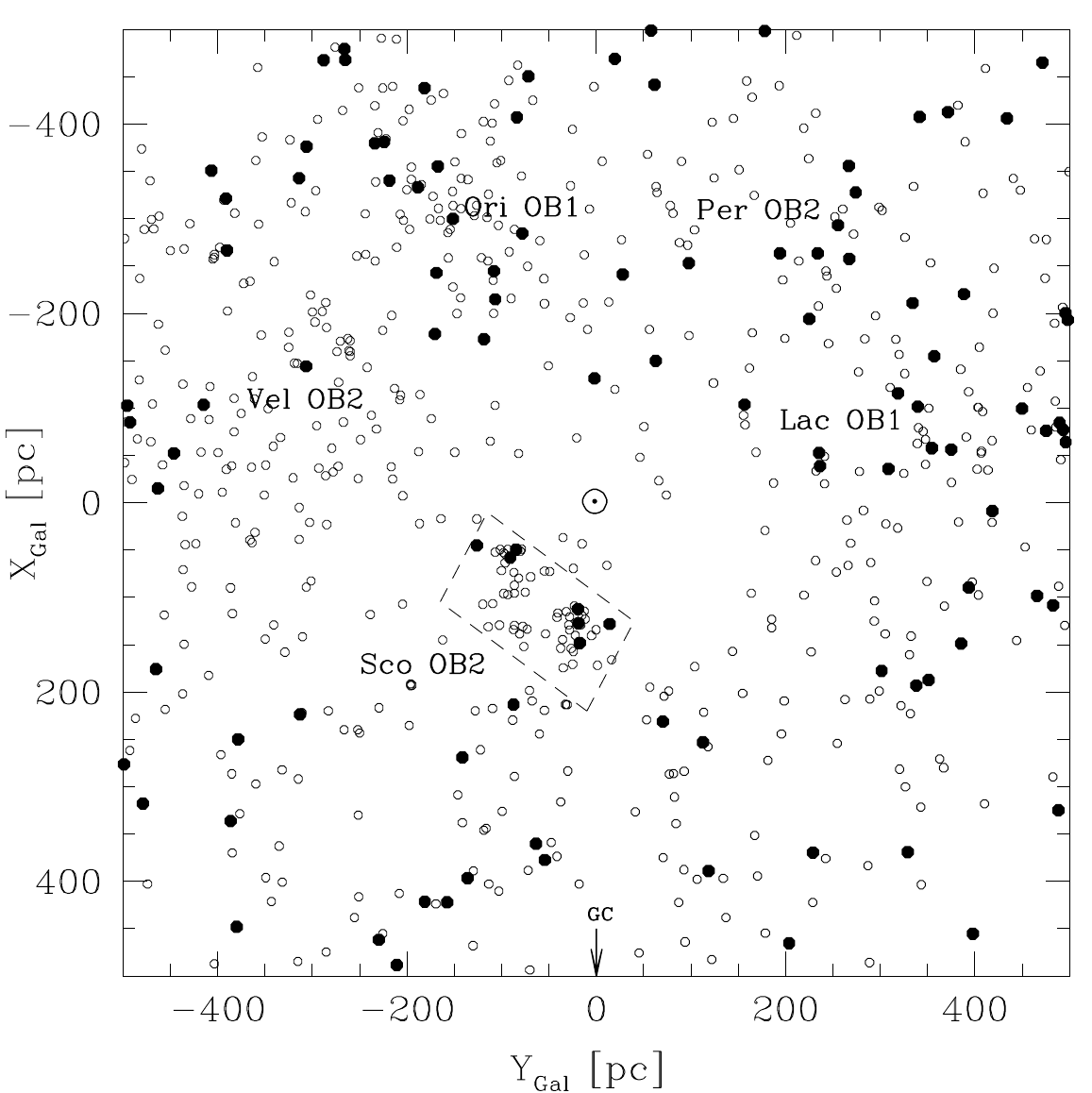}
  \caption{OB associations near the Sun, with Sco-Cen stars marked \citep[from][]{2008hsf2.book..235P}. This top view onto the Galaxy in the local vicinity shows that several nearby groups of massive stars can be identified, with the Scorpius-Centaurus group as the closest to the Sun. This compilation is based on Hipparcos position measurements of nearby stars \citep[for details see][]{2008hsf2.book..235P}.}
   \label{FigOBassoc}%
   \end{figure}

Massive stars are key agents shaping the interstellar medium, through ionizing starlight, stellar winds and supernovae. This has been recognized throughout galaxies \citep{2004ARA&A..42..211E}, and also locally in the surroundings of our Sun \citep{1995SSRv...72..499F, 1992A&A...262..258D}. Massive stars tend to be located in clusters \citep{2003ARA&A..41...57L,2007ARA&A..45..481Z}. Stellar winds and explosions are expected to create interacting gas shells \citep{1977ApJ...218..377W}, merging of hot cavities may form superbubbles \citep{1988ARA&A..26..145T, 1988ApJ...324..776M, 2008NewA...13..163C}. The environment of the Sun shows clear signatures of massive-star action, though located in a seemingly quiet region of the Galaxy, about 8~kpc from its center and about 4--5~kpc from the molecular ring identified as the region of the Galaxy's highest star-forming activity \citep{1984SciAm.250...42S}. The interstellar medium near the Sun is characterized by a rather complex morphology \citep{1995SSRv...72..499F}. The Local Bubble, a rather tenuous cavity, surrounds the Sun \citep{1998LNP...506.....B}. Several loops/shells reminiscent of supernova remnants have been identified \citep{1992A&A...262..258D}. The Local Bubble is the most-nearby of such cavities, and finding its origins is a prominent current science topic \citep{2006A&A...452L...1B, 2009Ap&SS.323....1W}. Groups of massive stars are found within a distance of several hundred parsec \citep{1999AJ....117..354D}, and also nearby clouds with current star forming activity as evident in T Tauri stars, HII regions, and embedded IR sources. 
The stellar association of Scorpius-Centaurus (Sco OB2) constitutes a prominent group of most-nearby stars (see Fig.~\ref{FigOBassoc}). It is located  at a distance of about 100--150~pc \citep{1992A&A...262..258D, 1999AJ....117..354D, 1999AJ....117.2381P}. Apparently it plays a major role for the nearby interstellar medium and accounts for a significant fraction of the massive stars in the solar vicinity \citep{1999AJ....117..354D}. This association shows several subgroups of different ages (estimated as 5, 16, and 17~My, with typical age uncertainties of 2~My \citep{1989A&A...216...44D} (see however \citet{2008ApJ...688..377S, 2009SSRv..143..437F}). Its location in the sky is centered $\sim$20$^{\circ}$  above the plane of the Galaxy and hence sufficiently-distinct, unlike other Galactic source groups which would be superimposed along the line of sight towards the Galactic ridge or molecular ring, for example. Our target is \Al radioactivity from this region. 

Massive stars are believed to form from dense parts of giant molecular clouds, and have typical evolution time scales of 1--10~My \citep[see reviews by ][]{2007ARA&A..45..565M,2007ARA&A..45..481Z}. 
One of the most important questions in models of the star formation process is the onset of release of energy and matter by the most-massive stars and their effects on the surrounding clouds \citep{2007astro.ph..3036O,2008MNRAS.386....3C}. Such feedback on star formation in dense clouds can be either negative (termination of star formation through dispersal of the natal cloud) or positive (triggering of further star formation by compression of cloud material). The massive-star clusters are believed to be dispersed within a typical time scale of 10~My \citep[e.g.][]{2009A&A...498L..37P}.  

Within a group of massive stars, Wolf-Rayet-phase winds carry the products of core hydrogen burning including \Al and lead to a steeply-rising \Al content in surrounding ISM about 3 My after those stars have been formed \citep{2009A&A...504..531V}.  Interstellar \Al content then decreases again within 1--2~My, from radioactive decay and because the less-massive, more slowly-evolving stars do not develop the Wolf-Rayet phase below M$_{i}\sim$25 \Msol.  Trailing this Wolf-Rayet-dominated initial \Al production, contributions from core-collapse supernovae take over to enrich the ISM with \Al.  This production smoothly declines as less and less-massive stars terminate their evolution in a supernova, and the interstellar \Al content fades after $\sim$~7--8~My with a tail extending to $\sim$20 My. In general, stellar groups with an age in the range of 3--20~My may be considered plausible sources of interstellar \Al  \citep[see][and references therein]{2009A&A...504..531V}. The youngest stellar subgroup of the Sco-Cen association is the Upper Scorpius group at an age of about 5~My. This should correspond to the peak of the expected \Al content \citep[see Fig.~2 of ][]{2009A&A...504..531V}.

\Al is ejected through stellar winds and supernova explosions at typical velocities around 1000~km~s$^{-1 }$\citep{2002AcA....52...81N, 2000ApJ...531L.123K}, which will be slowed down by interactions with circumstellar material, and by the reverse shock which results from the discontinuity between the expanding ejecta and ambient ISM.  \Al decays with an exponential lifetime of 1 My. Hence  \Al decay occurs along the ejecta trajectory in the surrounding ISM. Most likely that will be hot and probably turbulent, from the injected radiation, wind and supernova energies. Upon $\beta^+$-decay, a $\gamma$-ray line at 1808.63~keV is emitted, together with a positron \citep{1990NuPhA.521....1E}. 
High-resolution spectroscopy of the \Al $\gamma$-ray line contributes to test massive-star nucleosynthesis, and also the general scenario of their formation in groups and sequences in an independent way, through such a radioactivity clock. Also, the \Al $\gamma$-ray line centroid and width reflect the kinematics of massive-star ejecta at late times: The Doppler effect will encode bulk motion and turbulence in the observed \Al $\gamma$-ray line profile. 
Here we report our search for \Al $\gamma$-ray emission from the nearby Sco-Cen association and its subgroups with the Compton Observatory and INTEGRAL $\gamma$-ray instruments.

  \begin{figure}   
   \centering
  \includegraphics[width=0.48\textwidth]{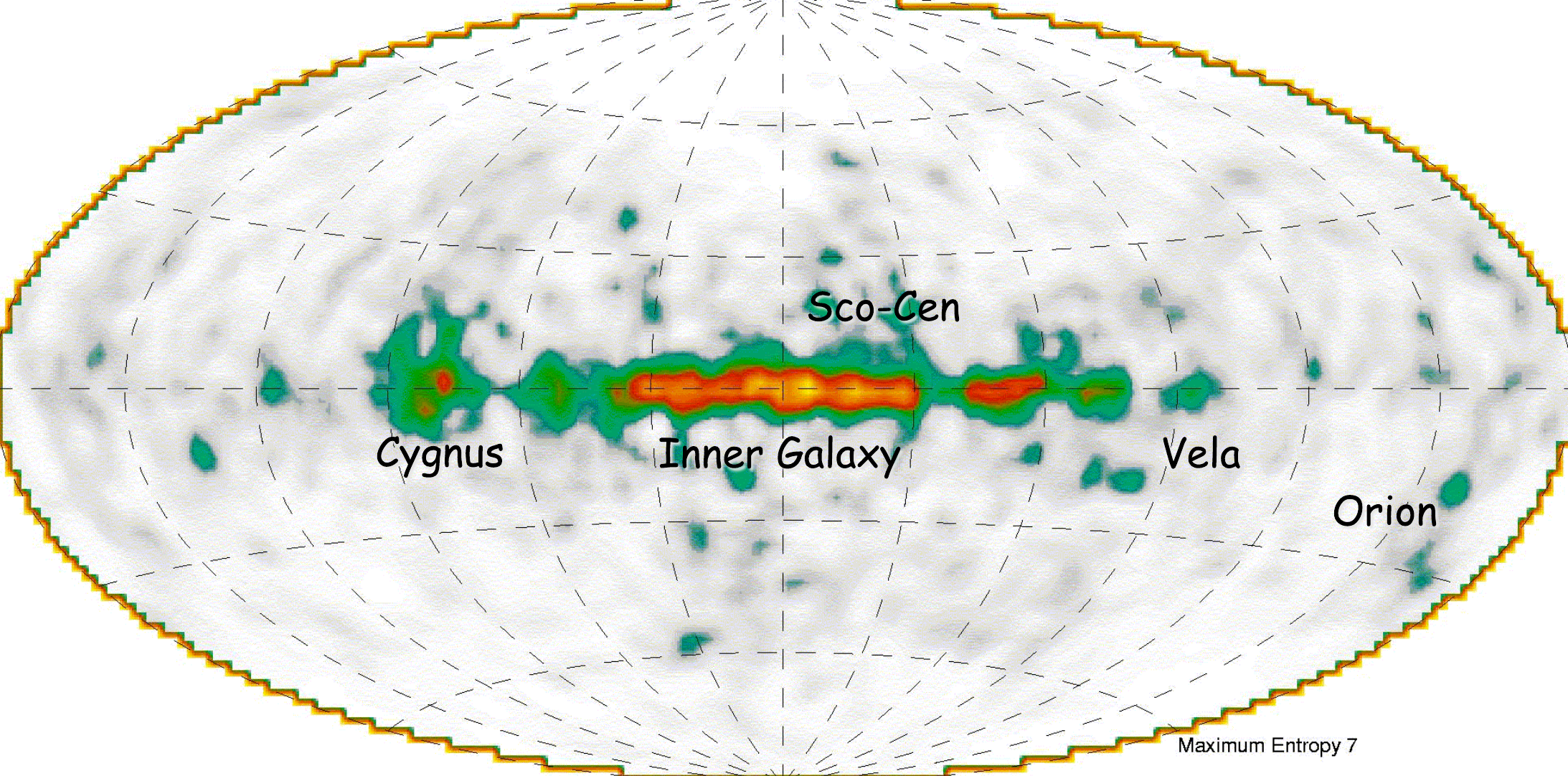} 
  \includegraphics[width=0.3\textwidth]{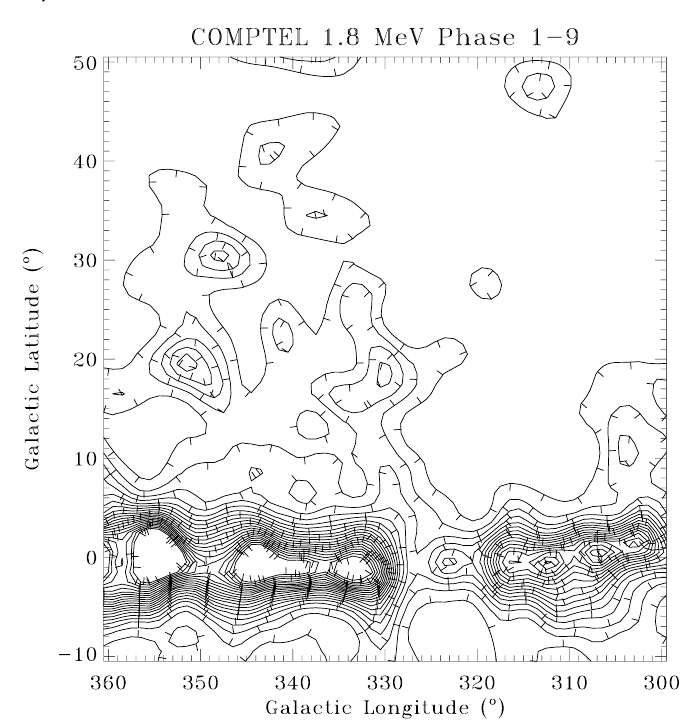}
   \caption{All-sky image ({\it top}) and expanded view towards Sco-Cen ({\it bottom}) of \Al $\gamma$-rays as derived from COMPTEL observations \citep{2001ESASP.459...55P}}
   \label{FigCOMPTEL-Maps}%
   \end{figure}

\section{Earlier Data and Hints}

The COMPTEL instrument was operated aboard the Compton Gamma-Ray Observatory mission of NASA through 1991--2000 \citep{1993SciAm.269...68G,1993ApJS...86..657S}. It was a Compton telescope, detecting $\gamma$-rays in the MeV range through a correlated detection of a Compton scatter process in its upper detector plane and the secondary photon from this scatter process being absorbed in its lower detector plane  \citep[see][for a detailed instrument description]{1993ApJS...86..657S}. With its large field of view of about 30~\degree, COMPTEL performed a sky survey for MeV $\gamma$-ray lines. It thus provided the first all-sky maps of \Al emission \citep{1995A&A...298..445D}, which showed the \Al emission extending along the entire plane of the Galaxy. From these data, it was concluded that massive stars are the dominating sources of \Al ejection into the Galaxy's interstellar medium, through Wolf-Rayet winds and core-collapse supernovae \citep[see][for a review]{1996PhR...267....1P}.

Tantalizing hints for \Al emission from a region above the plane of the Galaxy and reminiscent of the Sco-Cen region was noticed in COMPTEL's \Al all-sky image as derived from Maximum-Entropy deconvolution of the data (Fig.~\ref{FigCOMPTEL-Maps}, \citet{1995A&A...298..445D, 1998LNP...506..389K, 2001ESASP.459...55P}). The significance of this result is low, however, and depends upon the specific formulation of the hypothesis. 
A maximum-likelihood test for the emission in the Sco-Cen region obtained 2.1~$\sigma$ significance only, from separately modeling the data with the deconvolved map parts of the region of interest (defined as a circle of 15~degrees radius around (l,b)=(340\degree\ ,20\degree)), this representation of the Sco-Cen region led to only marginal improvement in the maximum-likelihood fit to 9 years of observation above the remainder of the MaximumEntropy map. Similar values could be obtained for other intermediate-latitude regions of the sky of similar extent, and thus the signal was not pursued any further.

INTEGRAL, ESA's  INTernational Gamma-Ray Astrophysics Laboratory, has been operating in space since November 2002 \citep{2003A&A...411L...1W}. Its $\gamma$-ray spectrometer SPI \citep{2003A&A...411L..63V, 2003A&A...411L..91R} features an excellent spectral resolution of 3~keV at  $\sim$1800~keV, which compares to about 200~ keV spectral resolution for COMPTEL. Spatial resolutions of these instruments are similar, 2.7$^{\circ}$ for SPI and 3.4$^{\circ}$ for COMPTEL, with COMPTEL imaging being achieved through kinematical constraints from measuring the Compton scattering within the instrument, while SPI's imaging approach is through a coded mask casting a shadow onto its pixelated Ge-detector plane.  INTEGRAL's sky exposure has emphasis in the inner Galactic plane and some special extragalactic source regions. \Al emission from the sky was detected at the intensities expected from COMPTEL's earlier survey \citep{2003A&A...411L.451D}. Making use of the advance in spectral resolution, the line intensity and shape were determined for the inner Galaxy region, and could be used to determine the global properties of \Al in the Galaxy's interstellar medium \citep{2006A&A...449.1025D, 2006Natur.439...45D}. The $\gamma$-ray line turned out to be rather narrow and not significantly Doppler-broadened, consistent with the \Al-enriched interstellar medium being moderately turbulent at velocities in or below the 100~km~s$^{-1}$ range.  The Doppler shift of the $\gamma$-ray line centroid from large-scale rotation of the Galaxy also was indicated in these high-resolution data. This confirmed the presence of \Al throughout the plane of the Galaxy. The Galactic \Al content is determined from the integrated intensity of the $\gamma$-ray line to range between 2 and 3~\Msol, when plausible models for the spatial distribution of \Al in the Galaxy are employed to resolve the distance ambiguities along the lines of sight \citep{2006Natur.439...45D}. 

The spatial source distribution models used in these determinations all assume that  the \Al sources are adequately represented to first order by spiral-arm or exponential-disk descriptions of the Galaxy. The overall symmetry of the Galactic \Al emission as observed in the bright inner ridge in the first and fourth quadrants of the Galaxy is reassuring for such large-scale interpretation \citep{2009A&A...496..713W}. Only the Cygnus region's \Al emission appeared as a clear and prominent deviation from this view \citep{1996A&A...315..237D, 2002A&A...390..945K}, while variations of \Al emission along the inner ridge of the Galaxy as seen in the COMPTEL image were assumed to be due to the clumpiness of massive-star locations throughout the Galaxy (i.e., the ``beads-on-a-string'' seen in UV images of nearby galaxies when viewed face-on). Tests for local emission from nearby and special groups of OB associations such as attributed to the Gould Belt \citep{1997FCPh...18....1P, 2009MNRAS.397....2E} had been made on COMPTEL data \citep{1998LNP...506..389K}. Only marginal hints were found, and thus the large-scale interpretation of observed \Al was held up. 

But when nearby \Al sources,  such as may be expected from the Sco-Cen groups of stars as close as 100--150~pc  (see Fig. \ref{FigOBassoc}, and \citet{1964ARA&A...2..213B}) are superimposed on our viewing direction towards the Galaxy's inner ridge, a significant bias might occur. This could lead to an over-estimate of the Galaxy's \Al content, and of the scale height of the \Al sources in the Galactic plane, which is determined from \Al $\gamma$-ray data as about 130~pc \citep{2009A&A...496..713W}. It has been shown that the Galactic \Al content (and hence the nucleosynthesis yields, or supernova and star formation rates derived from it) could be lower by 30\% or more, if foreground emission from specific massive-star clusters would bias our inferred scale height towards high values \citep{2009A&A...506..703M}. Thus the real \Al content may be on the low side of the uncertainty range quoted before \citep{2006Natur.439...45D}. 
We exploit deeper observations made with INTEGRAL to follow up on these questions.

\section{SPI Observations and Data Analysis}

Continued SPI observations of the inner ridge of the Galaxy have reached a depth where spatial separation of the \Al emission begins to be possible. This allows spatially-resolved spectroscopy of the \Al $\gamma$-ray line, towards determination of \Al-enriched ISM characteristics for different regions of the Galaxy \citep[see][]{2009A&A...496..713W}. 
We use the same data  as analyzed by \citet{2009A&A...496..713W} to determine possible \Al emission from massive star nucleosynthesis in the Scorpius-Centaurus region. 
These data include orbits 43--623 from 5~years of INTEGRAL observations, with a total exposure of 61~Ms. 
We analyze single-detector hits (SE)  in an energy range embracing the \Al line at 1808.63~keV, specifically 1785--1826~keV,  in 0.5~keV bins. 
Results are derived through fitting the set of binned event spectra by the composite of forward-folded sky signal and background models, allowing for intensity adjustments of each of these components in each energy bin. 
Spectra are obtained for different regions of the sky, as the adopted spatial distributions for each region are fitted separately, though simultaneously, to the complete ensemble of observational data. The subdivision of the sky into different sky regions of interest is limited by the additional parameters introduced into the fit. We minimize these here by representing the sky with two components only, one for the large-scale Galactic emission, and one for the region of interest. Varying the region of interest, we analyze the quality of fits obtained, possible systematics and background artifacts, and the consistency of our results.

  \begin{figure}   
   \centering
   \includegraphics[width=0.48\textwidth]{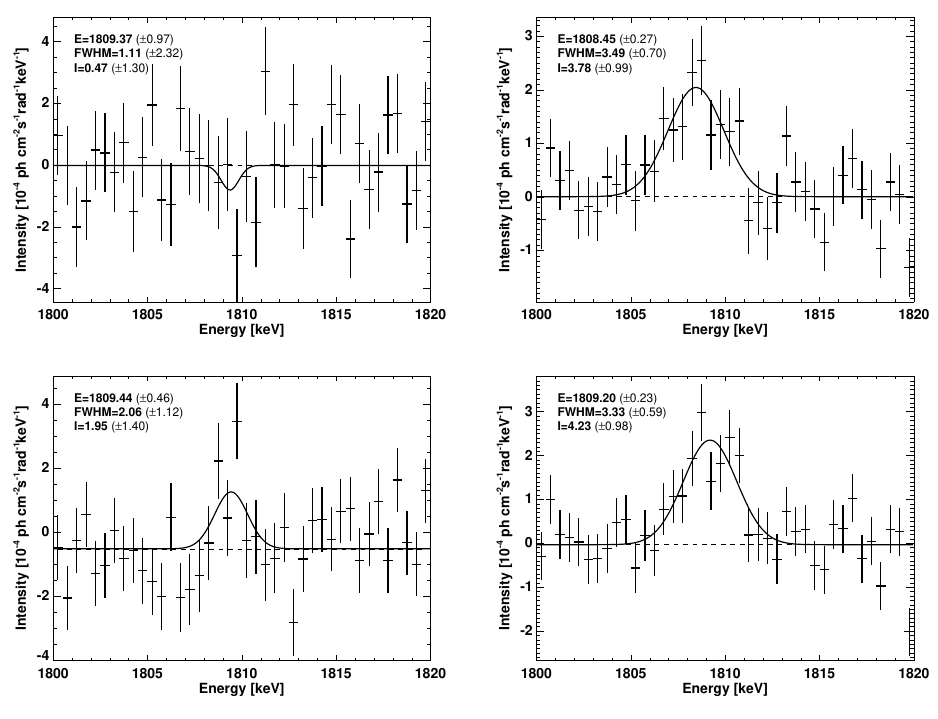}
   \caption{The $\gamma$-ray spectrum obtained from fitting a homogeneously-bright sky to to the first ({\it top}) and fourth ({\it bottom}) quadrants of the Galaxy. The {\it lefthand} figures show emission at intermediate positive latitudes 5$^{\circ}$--20$^{\circ}$, the  {\it righthand} figures show emission from the bright Galactic disk in the latitude interval [-5$^{\circ}$,5$^{\circ}$] \citep{2007PhDTUM...Wang} . }
   \label{FigSpectrumSPI_WW}%
  \end{figure}

As a guide towards our study of Sco-Cen emission, we first study separating Galactic-plane \Al emission from emission at intermediate latitudes,. We find significant excess of \Al emission in the northern part of the Galactic plane (Fig.~\ref{FigSpectrumSPI_WW}, lower-left graph). Here, an exponential disk has been divided in latitude at 5$^{\circ}$ and separately fitted in intensity. This indicates that in the fourth quadrant of the Galaxy, there is excess line emission at intermediate latitudes, if compared to the first Galactic quadrant. 
We refine this through restricting the extra emission component to the Sco-Cen region only, while the second sky emission model component includes the remainder of the sky.

As we include all data for best constraints on instrumental background, we also require an adequate model of the full sky in \Al. We plausibly employ several alternatives: The \Al sky as observed with COMPTEL and determined with Maximum-Entropy or Multi-Resolution Expectation Maximization algorithms for imaging deconvolution \citep{1999A&A...345..813K}, but also geometrical models which had been found to adequately represent \Al emission, from COMPTEL and SPI analysis \citep[e.g.][]{1999A&A...344...68K, 2006A&A...449.1025D}. Examples of the latter are double-exponential disks with Galactocentric scale radius 3.5 to 4~kpc and scale heights 130 to 300~pc, the free-electron model derived from pulsar dispersion measurements by \citet{1993ApJ...411..674T}  (we use the NE2001 model, \citet{2002astro.ph..7156C}, with an imprinted vertical scale height of the exponential distribution above the plane of 180 or 140~pc), or the FIR emission from warm dust as measured with COBE/DIRBE across the sky \citep{1992ApJ...396L...7B}.  

The above models based on non-$\gamma$-ray observations may not represent nearby \Al from Sco-Cen: They characterize the large-scale Galactic \Al contributions, because they trace different massive-star related phenomena with indirect connection to \Al per invividual source. On the other hand, the COMPTEL map would obviously include \Al from Sco-Cen, yet is mediated by COMPTEL's instrumental response and the deconvolution algorithm to derive the map, and thus not reliable in this detail. We therefore prefer to use the COMPTEL \Al map for the general large-scale \Al emission, and here adopt its MREM variant \citep{1999A&A...345..813K}. We cut out from it a region which lies above the plane of the Galaxy and towards the Sco-Cen sources here, defined as a circle centered at (l,b)=(350$^{\circ}$ ,20$^{\circ}$) with a radius of 8$^{\circ}$.  \Al is ejected at velocities of $\sim$1000--1500~km~s$^{-1}$ and slows down in interstellar medium from interactions with the gas; in our large-scale Galactic average, but also in the Cygnus region, we find velocities are certainly below 200 km~s$^{-1}$ from Doppler broadening limits on the observed line width \citep{2006A&A...449.1025D,2009A&A...506..703M}. The size of our region of choice corresponds to \Al coasting with interstellar velocities of 10~km~s$^{-1}$, hence will not include faster \Al and miss part of the production in this region. We chose this conservatively-small region to avoid misinterpreting \Al $\gamma$-rays overlapping directions towards the Galaxy's disk. We test spatially-unstructured models representing the Sco-Cen region, such as a homogeneous-intensity circle centered at this location, or an exponential disk split into a part below 5$^{\circ}$ latitude containing the bulk emission and above 5$^{\circ}$ latitude containing emission from Sco-Cen, or the extracted part of a COMPTEL \Al map suppressing small-scale structure. Detection significances vary between 5 and 7~$\sigma$ for the cases studied, derived fluxes typically vary by $\leq$10\% for different spatial models of the Sco-Cen region itself; this is well within statistical uncertainties.  

The instrumental background created by cosmic-ray interactions with the spacecraft material dominates the measured counts, and only about 1\% of the total counts is contributed by the signal from the sky. Therefore, background modeling is crucial, and artifacts from inadequacies of a background model may occur.  We therefore also test for \Al emission in other regions of the sky, where no such emission is plausibly expected. For most-representative tests, we chose regions of similar exposure and similar offset from the bright \Al emission along the inner Galactic ridge, i.e., the region south of the plane of the Galaxy at longitude 350$^{\circ}$ and latitude -20$^{\circ}$, and a region in the first quadrant of the Galaxy at (l,b)=(10$^{\circ}$,20$^{\circ}$). 

The characteristics of celestial \Al emission, i.e. bulk motion and turbulence within the source region, translate into Doppler broadening and shift of the observed $\gamma$-ray line. SPI Ge detectors show an intrinsic energy resolution at 1809~keV of 3~keV (FWHM), variations of the energy calibration remain below 0.05~keV \citep{2004ESASP.552..713L}. Effects degrading the detector resolution with time due to cosmic-ray irradiation in orbit, and the periodic annealings which successfully restore spectral resolution, are all relevant for strong signal where line shape details matter, and can be neglected for our rather weak target. Here we determine line centroids and widths from Gaussian fits, and find using our more-accurate asymmetric response including degradation effects does not affects our findings. Our spectral result parameters are limited by statistical precision, and overall systematic uncertainties (from energy calibration stability and from using a Gaussian spectral response) are estimated as 0.1~keV.

\section{Results}

  \begin{figure}   
   \centering
   \includegraphics[width=0.48\textwidth]{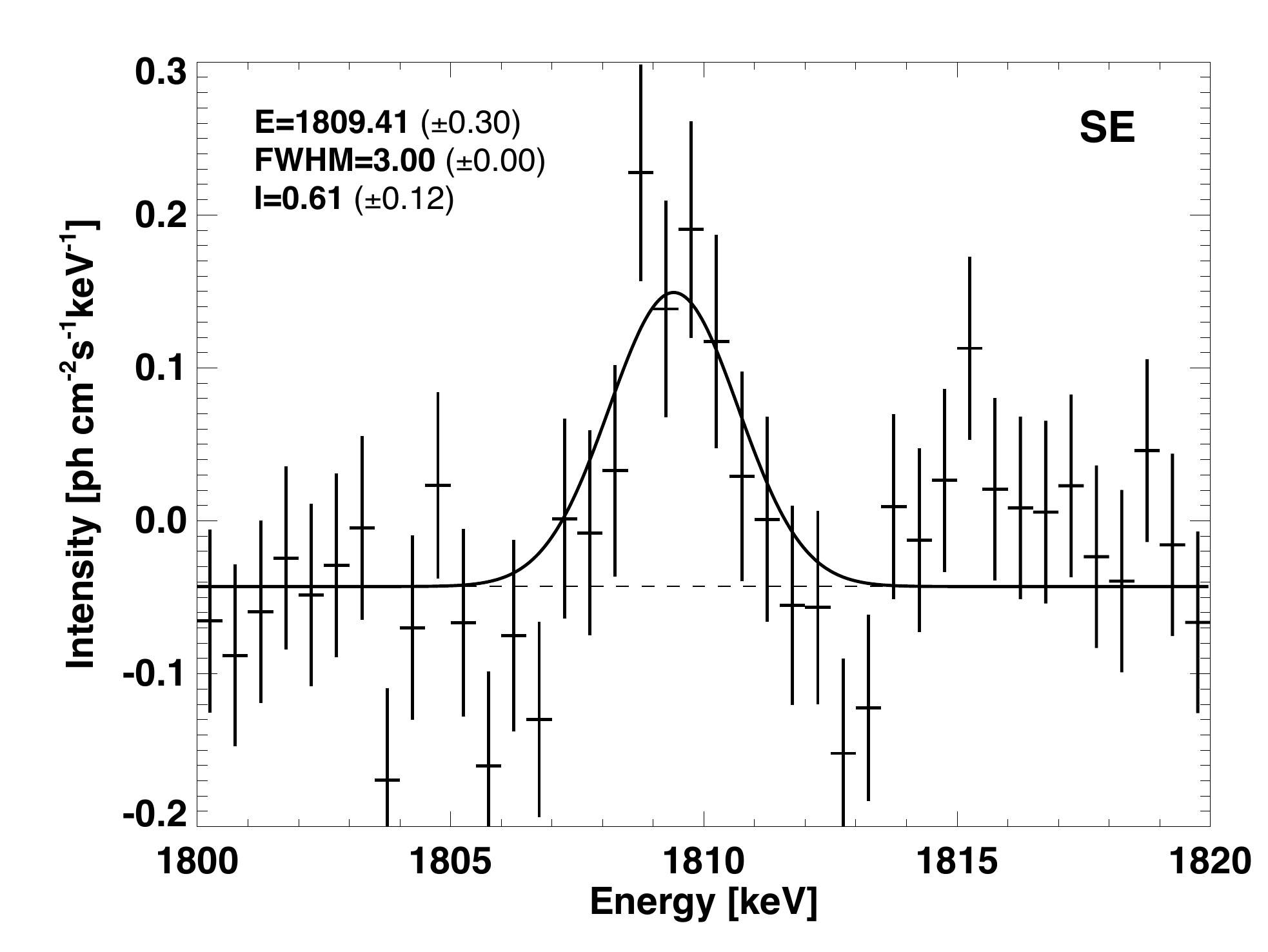}
   \caption{The $\gamma$-ray spectrum of the Sco-Cen region. This was obtained from fitting a 10$^{\circ}$ diameter patch of the sky representing candidate \Al sources of this region together with a large-scale Galactic-emission model (here: MREM \Al image as derived from COMPTEL observations) in narrow energy bins to SPI observation data.}
   \label{FigSpectrumSPI_ML}%
  \end{figure}

Using the COMPTEL \Al map for this major portion of the sky, and an un-structured circular patch of the sky in the Sco-Cen region ((l,b)=(350$^{\circ}$,20$^{\circ}$) with a radius of 10$^{\circ}$), we obtain the Sco-Cen region spectrum shown in Fig.~\ref{FigSpectrumSPI_ML}. The \Al line appears in the spectrum, now much clearer and at a statistical significance of 6~$\sigma$ (from the likelihood ratio of the fit). The Sco-Cen signal robustly is found in all above-mentioned alternative spatial-distribution models for the large-scale emission.

\begin{figure}   
   \centering
   \includegraphics[width=0.48\textwidth]{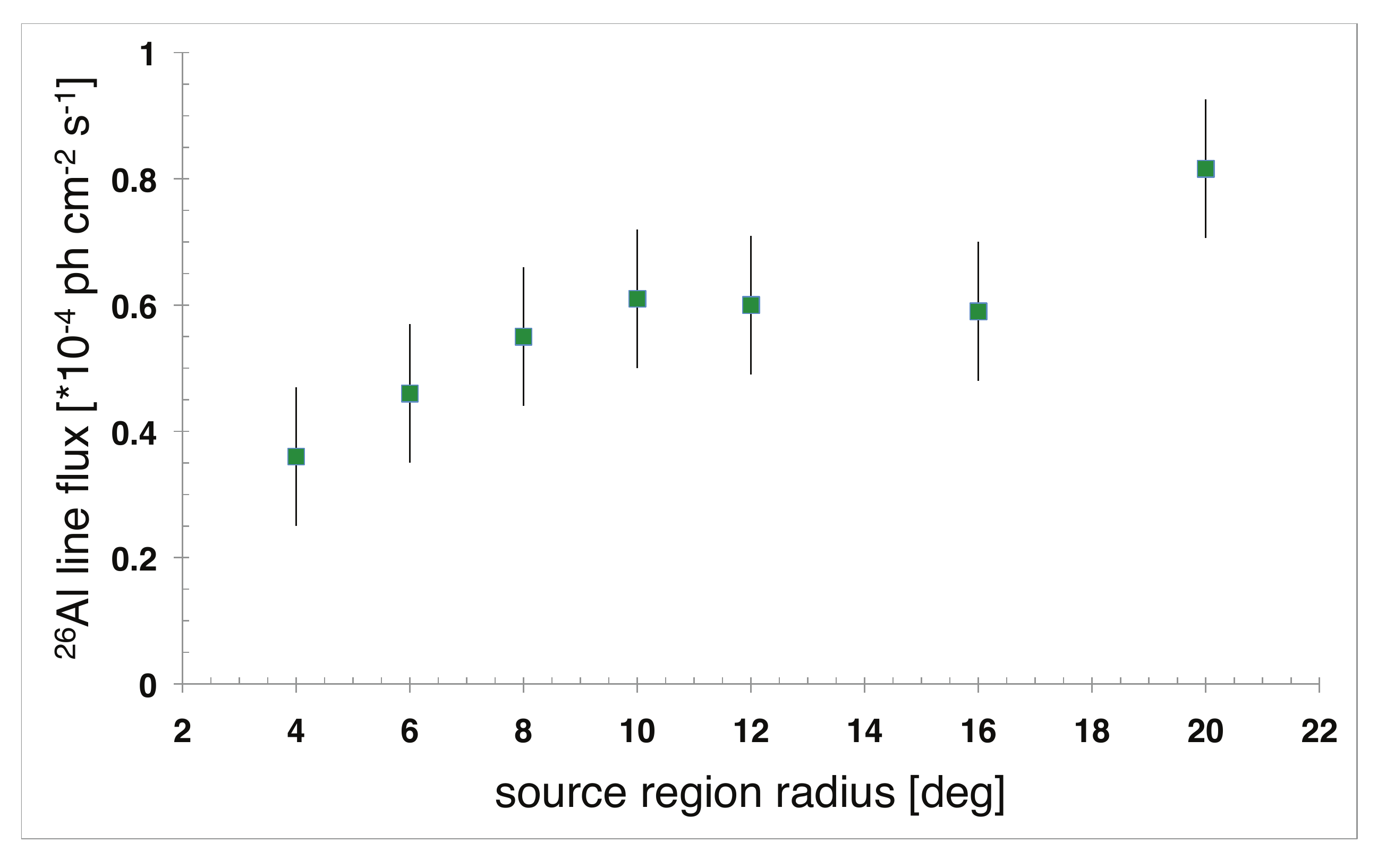}
   \caption{\Al flux increase with radius of our region of interest, centered on l=-10\degree/b=20\degree.}
   \label{FigSco_cumulative-flux}%
  \end{figure}

This \Al flux attributed to the Scorpius-Centaurus stars was estimated through variations of the candidate source region: We increase the circular region-of-interest radius from 4$^{\circ}$ to 20$^{\circ}$, deriving \Al line fluxes through 2-component spatial model fits. Fig.~\ref{FigSco_cumulative-flux} shows that the detected signal increases up to 10\degree radius, remaining constant for larger radii until the Galactic-plane emission is included (for the 20\degree case). None of our three reference regions shows such accumulation of signal with region size, their flux values fluctuate within uncertainties.

\begin{figure}   
   \centering
   \includegraphics[width=0.48\textwidth]{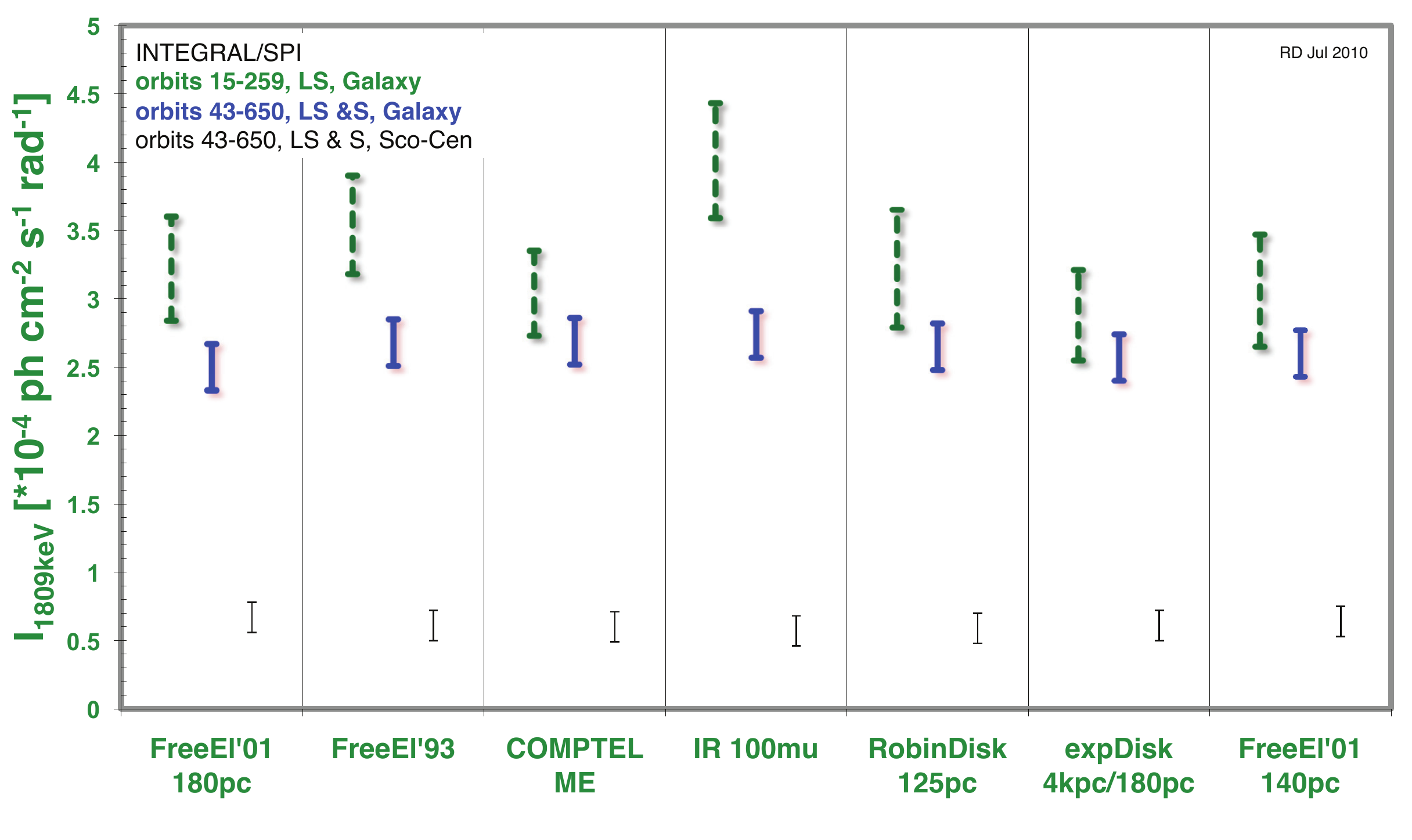}
   \caption{\Al flux determined for different plausible large-scale model distributions (details see text).  }
   \label{Fig_InnerGalaxyFlux}%
  \end{figure}

Including an emission component for this region attributed to nearby Scorpius-Centaurus stars, the fitted large-scale \Al flux from the bright ridge along the inner Galaxy  obtains lower values. In Fig.~\ref{Fig_InnerGalaxyFlux} we show a set of plausible models for the distribution of large-scale emission as discussed in \citet{2006Natur.439...45D}, as its flux values change through inclusion of the Scorpius-Centaurus component. We compare models based on free electrons, on infrared emission from dust, two analytical descriptions of the disk of the Galaxy, and the COMPTEL \Al gamma-ray image \citep[see details and references as described in][ appendix]{2006Natur.439...45D}.
Separating the Scorpius-Centaurus emission component (shown as thin symbols), the inner-Galaxy flux values (thick symbols; we use [-30\degree$<$l$<$30\degree],  [-10\degree$<$b$<$10\degree] approximating the region often referred to as ``the inner radian'') vary hardly among models. Accounting for observed data with these same large-scale model alone \citep[in an earlier analysis, ][ appendix]{2006Natur.439...45D},  variations among models were evident.  
Separating \Al emission from this foreground region,  the fitted large-scale Galactic \Al intensity   decreases by 19\% to 2.63~10$^{-4}$~ph~cm$^{-2}$s$^{-1}$rad$^{-1}$, with a statistical uncertainty of $\pm$0.2~10$^{-4}$~ph~cm$^{-2}$s$^{-1}$rad$^{-1}$ from this analysis using about twice as much data. 
Also shown in Fig.~\ref{Fig_InnerGalaxyFlux} are the fluxes we derive for the 8$^\circ$ Scorpius-Centaurus region source together with each of these alternate models. Variations are small and within statistical uncertainties.

We thus derive an \Al flux from our Scorpius-Centaurus region source of (6.1$\pm$0.11)~10$^{-5}$ph~cm$^{-2}$s$^{-1}$.  Flux uncertainties are  1.2~10$^{-5}$~ph~cm$^{-2}$s$^{-1}$(statistical)  and 1.0~10$^{-5}$~ph~cm$^{-2}$s$^{-1}$ (estimated systematics).  
(The variance of 0.2\% from the different large-scale models provides an estimate of systematic uncertainties from this aspect. 
Main systematic uncertainty is derived from our tests for a signal in the mirror-symmetric regions at (l,b)=(-10$^{\circ}$,-20$^{\circ}$) and (l,b)=(+10$^{\circ}$,+20$^{\circ}$). No significant celestial intensity is found for any such alternative source, but the scatter of derived fluxes with different candidate source-region sizes results in this estimate of 1.0~10$^{-5}$~ph~cm$^{-2}$s$^{-1}$. In these test we note that as now the Sco-Cen region is not part of the sky model, our extra component south of the plane obtains a negative-amplitude line at the position of the \Al line, which probably compensates for the now-too-intense large-scale model as fitted to our data.)
 
This \Al line flux from our Scorpius-Centaurus region source can be converted to an observed amount of \Al, adopting a specific distance for the source; for a distance of 150~pc we obtain 1.1~10$^{-4}$~\Msol of \Al.   

We note that an independent all-sky analysis on a somewhat smaller dataset represented the spatial distribution of Sco-Cen emission through a 2-dimensional Gaussian and obtained (6.2$\pm$1.6)~10$^{-5}$~ph~cm$^{-2}$s$^{-1}$ \citep{2009A&A...506..703M}. 

The \Al line from our Scorpius-Centaurus region source is found with a slightly  blue-shifted energy: From fitting the line shape with an asymmetric line shape as expected from variable spectral-response degradation between annealings, we determine a centroid energy of 1809.46~keV ($\pm$0.48 keV). 
When we compare to the \Al-decay laboratory value for the $\gamma$-ray energy of 1808.63~keV, this implies a blue shift of $\sim$0.8~keV. If interpreted as  a kinematic Doppler effect from bulk motion, this corresponds to about (137$\pm$75)~km~s$^{-1}$. 
Our uncertainties on the line centroid are $\pm$0.4 keV statistical (for fixed-width Gaussians) and 0.2~keV systematic uncertainty; here systematic uncertainty is estimated from the line centroid difference between Gaussian and instrumental line-shape fits to the Sco-Cen signal. This translates into the quoted bulk velocity uncertainty of 75~km~s$^{-1}$.

In Fig.~\ref{FigSpectrumSPI_ML} and most of our analysis for weak \Al signals, we rather use a Gaussian shape with fixed instrumental resolution to represent the line shape, in order to stabilize the fit from using less free parameters. Fitting the width of the observed line in such a Gaussian, we find a value of 2.1 $\pm$ 0.5~keV (FWHM),  formally below but consistent with the instrumental line width of 3.0~keV \citep{2003A&A...411L..91R} (which was used in the Gaussian line fitted in Fig.~\ref{FigSpectrumSPI_ML}). 


  \begin{figure}   
   \centering
   \includegraphics[width=0.48\textwidth]{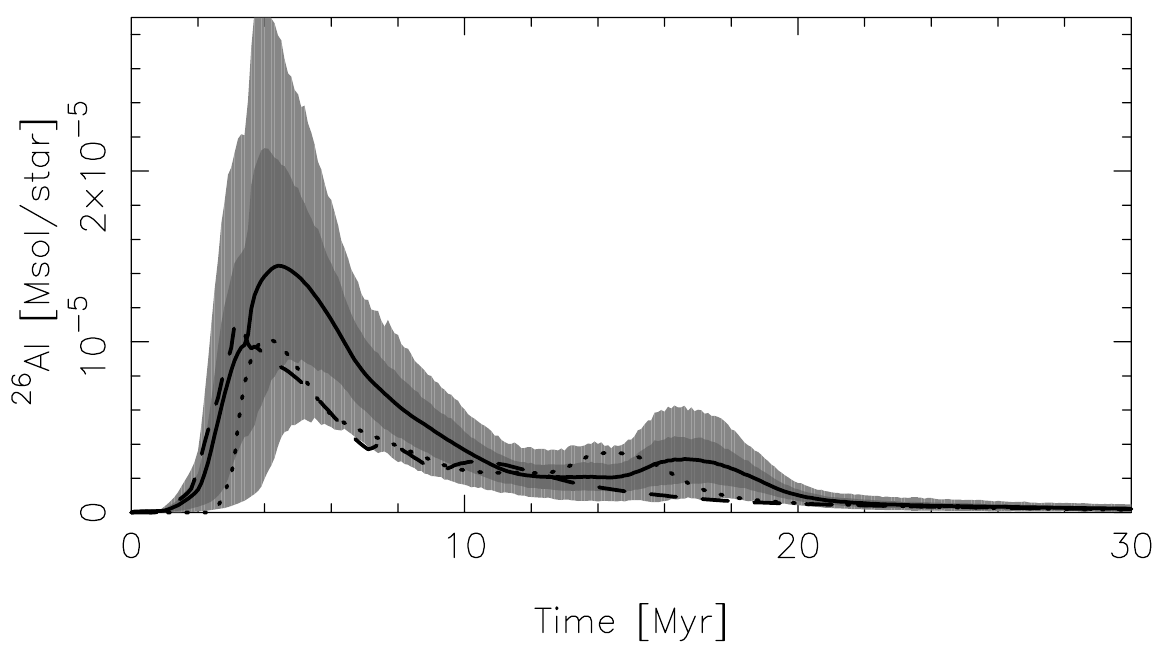}
   \caption{The \Al content of the ISM around a group of coeval massive-star groups, as it evolves with time \citep{2009A&A...504..531V}.}
   \label{FigAl_lightcurve}%
  \end{figure}

\section{Discussion}

The ``Scorpius-Centaurus''  complex of massive stars is the most-nearby region showing star formation within the last 10--20 My, with typical distances below 150 pc from the Sun\citep{1992A&A...262..258D, 1999AJ....117..354D, 1999AJ....117.2381P} . Our detection of \Al $\gamma$-rays confirms such recent and nearby massive-star activity. 
Since  \Al decay occurs with a time scale of 1~My, this implies that the massive stars providing the \Al sources were born less than $\sim$10~My ago (see Fig.~\ref{FigAl_lightcurve}).  
We observe an intensity in the $\gamma$-ray line at 1808.63~keV from \Al decay corresponding to a present \Al mass of   1.1~10$^{-4}$~\Msol  at 150~pc distance. The typical individual source-yield is about  1.1~10$^{-4}$~\Msol  for massive stars in the 8--40~\Msol range \citep[e.g.][]{2006ApJ...647..483L, 2007PhR...442..269W} \citep[see summary in appendix of][]{2006Natur.439...45D}, formally consistent with our derived \Al mass. The most massive star in Upper Sco was
presumably a $\sim 50\,M_\odot$ O5--O6 star, which may have exploded as a supernova about 1.5~Myr ago; the pulsar PSR~J1932+1059 may be its compact remnant \citep{2000ApJ...544L.133H,2004ApJ...604..339C}. But because of uncertainty of this event and no obvious supernova activity being evident in Sco-Cen less than 1~My ago (which would be seen e.g. as a supernova remnant), we are led to consider the observed \Al being a product of massive-star nucleosynthesis from a number of sources during an earlier epoch, rather than from a single event.

The \Al amounts ejected into the interstellar medium by a group of coeval massive stars evolves with stars developing Wolf-Rayet winds and finally exploding as supernovae (see Fig.~\ref{FigAl_lightcurve}, from our population synthesis model, \citet{2009A&A...504..531V}). 
From the number of stars in the Upper Sco group of stars (using 16 stars presently at masses M$>$8~\Msol, \citet{2002AJ....124..404P}) and an adopted age of 5~My, this population synthesis model predicts an intensity in the \Al $\gamma$-ray line of 7~10$^{-5}$~ph~cm$^{-2}$s$^{-1}$ from 1.2~10$^{-4}$~\Msol of \Al, which is consistent with our measurement. This would support age estimates derived from isochrone fitting to stellar groups (which are somewhat uncertain mainly from the calibration of pre-main-sequence models), and thus support Upper Sco as a region of most recent star formation, with a group age of about 5~My.

When we compare our observed \Al line centroid to the \Al-decay laboratory value for the $\gamma$-ray energy of 1808.63~keV, the evident blue shift of $\sim$0.8~keV corresponds to a kinematic Doppler effect from bulk motion at about (137$\pm$75)~km~s$^{-1}$. 
From our line width analysis we regard as unlikely additional astrophysical Doppler broadening from turbulent or isotropic motion at higher than normal interstellar gas velocities (i.e., $<$100~km~s$^{-1}$, which would correspond to 30\% astrophysical line broadening). 

      The cavity surrounding Upper Sco appears to expand with a velocity of about  10 ($\pm$ 2) km~s$^{-1}$, which, together with the relative motion of Sco-Cen stars with respect to the Sun of about 5 km~s$^{-1}$ implies an approaching relative velocity of the entire complex of few tens of km~s$^{-1}$ at most. Our measurement of a substantially-higher approaching velocity of \Al nuclei is incompatible with this. We would thus exclude \Al to decay while being associated with cavity walls: Decaying \Al moves at higher velocity. At the time of ejection, velocities of stellar wind or supernova material containing \Al would be about 1000--1500~km~s$^{-1}$. This velocity, if sustained over a \Al lifetime, would imply a propagation distance from the source of 1.3~kpc, and source ejecta would fill a sphere of that size, appearing isotropic from our position of the Sun (offset from the center of the sphere by only 0.1 its radius). But, with our analysis circle of 10\degree diameter around the source, we also impose a bias for \Al being near its source while decaying, sampling its effective streaming velocity over this volume sampled by our analysis. 
      
     Thus, our measured bulk velocity appears consistent with the small or absent \Al line broadening on one hand, and also with \Al being slowed down significantly from its high velocity at ejection, on a scale of 10~pc.  This appears consistent with our adopted scenario of observing stellar ejecta originating in the Sco-Cen stars and streaming into the pre-blown cavity around it, decelerating through turbulence and interactions with the cavity walls. 
      Other inferences of the morphology of interstellar gas around the Sun have independently suggested and supported this view \citep{1995SSRv...72..499F}. In particular, a streaming motion of interstellar gas from the general direction of Sco-Cen has been identified \citep[also referred to as the local interstellar wind (LISW); see, e.g.,][]{2003A&A...411..447L, 1995SSRv...72..499F}.    
      It also implies that our choice of a 10$^{\circ}$-sized region to identify \Al from Sco-Cen may miss a sigificant part of the emission, as the region filled with \Al by Sco-Cen should be more extended on the sky.  We cannot detect such more extended emission, both because it may be weaker than the main feature we report here, and because other locations of Scorpius-Centaurus stars overlap with Galactic-disk viewing directions. 
      We will pursue this aspect in a refined modeling approach based on details of the observed stellar population nearby  (Ohlendorf et al., in preparation).

Further exploring this question of how the nearby regions of the Galaxy have evolved, we may study the broader history of the Scorpius-Centaurus region:        
Stellar subgroups of different ages would result from a star forming region within a giant molecular cloud if the environmental effects of massive-star action of a first generation of stars (specifically shocks from winds and supernovae) would interact with nearby dense interstellar medium, leading to  propagating or triggered star formation. Later-generation ejecta would find the ISM pre-shaped by previous stellar generations. Such a scenario was proposed for the  Scorpius-Centaurus Association,   based on its different subgroups \citep{1989A&A...216...44D, 2002AJ....124..404P} and several stellar groups around it \citep[e.g.][]{2008hsf2.book..235P, 2008A&A...480..735F}. 
Indications of recent star formation have been found in the L1688 cloud as part of the $\rho$~Oph molecular cloud, and may have been triggered by the winds and supernovae which produced the \Al we observe. The young  $\rho$~Oph stars then could be interpreted as the latest signs of propagating star formation originally initiated from the oldest Sco-Cen subgroup in Upper Centaurus Lupus \citep{2008hsf2.book..351W}. 
The origin of runaway stars and nearby pulsars also provides hints towards a recent supernova in the solar vicinity \citep[see, e.g.][]{2000ApJ...544L.133H,2002AJ....124..404P}.

Many proposed scenarios of triggered star formation are only based on relatively weak evidence, such as the presence of Young Stellar Objects (YSOs) near shocks caused by massive stars.  
Positional evidence alone is not unequivocally considered to prove triggered star formation. Much more reliable conclusions can be drawn if the ages of the young stellar populations can be determined and compared to the moment in time at which an external shock from another star formation site arrived. Agreement of these timings would add important arguments. Radioactive clocks such as \Al can provide such information in this context in a most-direct way, being driven by physics which is rather independent of the environmental conditions of the dynamic ISM.
Therefore, we aim to constrain \Al emission from the presumably-older subgroups of the Sco-Cen association in a subsequent study (Ohlendorf et al., in preparation).
   
   Supernovae are also expected to have far-reaching implications such as a biologically-relevant UV flash and enhanced cosmic-ray intensity which may be relevant for climatic changes \citep[as discussed, e.g., in][]{2007SSRv..130..341L, 2007A&G....48a..18S}.  It is interesting that the discovery of $^{60}$Fe isotopes in Pacific ocean crust material on Earth \citep{2004PhRvL..93q1103K} provides strong evidence of a nearby supernova event about 2--3~My ago, which may be attributed to Sco-Cen stars as well.  It will remain exciting to obtain refinement in correlated investigations of observables from such nearby cosmic violence.    
   
We note that inclusion of nearby \Al from Scorpius-Centaurus stars in the determination of the large-scale \Al emission from the Galaxy reduces the large-scale flux value by 19\%. This translates directly into a corresponding reduction of the Galaxy's total \Al mass from 2.8 to 2.25 \Msol, propagating to inferences based on this \Al mass such as the core-collapse supernova rate and the interstellar isotopic ratio $^{26}$Al/$^{27}$Al  \citep[see][]{2006Natur.439...45D}. Note that our earlier analysis accounted for this aspect of large-scale source distribution uncertainty through quoting a systematic uncertainty of 30\%, and specifically for the Galaxy's \Al mass a value of (2.8$\pm$0.8)~\Msol. As shown recently by \citet{2009A&A...506..703M} with a ring model of 90~pc scale height for Galactic sources and specific model components for the Cygnus and Sco-Cen regions, the \Al mass in the Galaxy may well be even below 2~\Msol (they determine (1.7$\pm$0.2) \Msol) for this geometry; note that \citet{2009A&A...496..713W} find an \Al emission scale height of 130~pc). With now a clear detection of foreground \Al emission from the Scorpius-Centaurus stellar group(s), we are in a better position to establish improved 3-dimensional \Al source distribution models for our Galaxy to reduce the line-of-sight uncertainties in source distances, accounting for the roles of candidate stellar groups such as Cygnus and Scorpius-Centaurus, but also Gould Belt associations and associations in spiral-arm regions such as Carina. Together with reduced statistical uncertainty from more observational data, this will result into a more realistic  estimate of the Galaxy's \Al mass and derived quantities. Such a study is beyond the scope of this paper. 
   
 In conclusion, our observation of \Al from the Sco-Cen region strengthens evidence for nearby massive star activity in the recent past (about 20~My or less). With improved observations, we aim at spatially-resolved study of the \Al emission from this region, to discriminate among Sco-Cen subgroups and their \Al contributions, and to ad diagnostics of massive-star activity in the solar vicinity.

\begin{acknowledgements}
      Part of this work was supported by the Excellence Cluster "Origins and Evolution of the Universe".
      The SPI project has been completed under the responsibility and leadership of CNES/France.We are grateful to ASI, CEA, CNES, DLR, ESA, INTA, NASA and OSTC for support.
      The SPI anticoincidence system is supported by the German government through DLR grant 50.0G.9503.0.
      We thank the anonymous referee for constructive suggestions which helped to improve this paper.
\end{acknowledgements}

\bibliographystyle{aa}
%


\end{document}